\documentclass[journal, onecolumn]{new-aiaa}
\usepackage[utf8]{inputenc}
\usepackage{textcomp}

\usepackage{float}

\usepackage{graphicx}
\usepackage{amsmath}
\usepackage[version=4]{mhchem}
\usepackage{siunitx}
\usepackage{longtable,tabularx}
\usepackage[labelformat=simple]{subcaption}

\usepackage{caption}
\usepackage{subcaption} 

\usepackage{xcolor}
\usepackage{ulem}

\title{In Situ Imaging of Parachute Textile Micro-Mechanics Under Tensile Load}

\author{Cutler A. Phillippe\footnote{Ph.D Student, Aerospace Engineering, 104 S Wright St, Urbana, IL 61801, AIAA Student Member. Email: Cutlerp2@illinois.edu 
} and Francesco Panerai\footnote{Assistant Professor, Aerospace Engineering, 104 S Wright St, Urbana, IL 61801, AIAA Associate Fellow.}}
\affil{Department of Aerospace Engineering at The University of Illinois at Urbana-Champaign, Urbana, IL 61801}

\author{Laura Villafa\~{n}e Roca\footnote{Assistant Professor, Aerospace Engineering, 104 S Wright St, Urbana, IL 61801, AIAA Senior Member.}}
\affil{Department of Aerospace Engineering at The University of Illinois at Urbana-Champaign, Urbana, IL 61801}
\affil{Beckman Institute, University of Illinois at Urbana-Champaign, Urbana, IL 61801}

\begin{document}

\footnotetext{Preliminary work presented at AIAA SciTech 2024 in Orlando, FL, Jan 8-12, AIAA 2024-1003}

\maketitle
\begin{abstract}
Micro-mechanics of parachute fabrics under tensile loads is studied using in situ X-ray micro-tomography. Results are presented for two nylon textiles commonly used in parachute systems, MIL-C-7020H Type III and MIL-C-44378(GL) Type II. Textiles are subjected to incremental tension using a custom apparatus that loads the fabric radially, and the microstructure is imaged sequentially at steady load conditions. Micro-tomography images are processed using learning-aided segmentation and a custom processing pipeline that tracks the locations and morphological properties of individual tows on 3-D datasets. Results are used to reconstruct tow micro-scale properties and meso-scale strains. Our findings reveal a direct relation between the fabric architecture and the meso-scale mechanics. Warp tow pretensioning during manufacturing is found to affect decrimping and the anisotropy of the textile strains. Areal porosity increase with tension is quantified and a geometric model for pore opening under incremental load is validated.
\end{abstract}

\section*{Nomenclature}
{\renewcommand\arraystretch{1.0}
\noindent\begin{longtable*}{@{}l @{\quad=\quad} l@{}}
$A$  & tow cross-sectional area, $\mu$m$^2$ \\
$D$ & flexural modulus, Pa \\
$D_h$ & hydraulic diameter, $\mu$m \\
$d$, $w$ & tow projected diameter (width), $\mu$m \\
$F$ & force, N \\
$h$ & tow cross-section vertical thickness, $\mu$m \\
$r$ & principle radial axis from plunger center \\
$t$ & equivalent thickness of textile of equal solid volume, $\mu$m \\
$x$ & principle axis along weft tows \\
$y$ & principle axis along warp tows \\
$z$ & principle vertical axis normal to plunger and textile surface \\
$\beta$ & tow centroid-to-centroid spacing, $\mu$m \\
$\delta$ & plunger deflection, mm \\
$\epsilon_h$  & strain of tow cross sectional thickness, $\mu$m/$\mu$m \\
$\epsilon_w$  & strain of projected tow width, $\mu$m/$\mu$m \\
$\epsilon_x$  & strain of unit cell along principle $x$ axis, $\mu$m/$\mu$m \\
$\epsilon_y$  & strain of unit cell along principle $y$ axis, $\mu$m/$\mu$m \\
$\gamma$ & areal porosity, also known as geometric porosity, $\mu$m$^2$/$\mu$m$^2$ \\
$\nu$ & Poisson's ratio, $\epsilon$/$\epsilon$ \\
$\sigma_{rr}$ & radial stress boundary condition, MPa \\
$\phi$ & crimp angle about textile centerline, rad\\
\end{longtable*}}
\twocolumn
\section{Introduction}

Parachutes play a critical role in Entry, Descent, and Landing (EDL) systems~\cite{1,2,3}, enabling safe deceleration of payloads and crew during planetary landings on bodies harboring an atmosphere. New parachute designs are often modifications of heritage systems and heavily rely on costly design cycles of sub and full-scale model testing to evaluate performance. Modifications to the textile weaves are generally informed by the rated properties of previous textiles, including permeability, tensile testing data, and other manufacturing specifications~\cite{1,2,3,4,7020H,44378,6}. The availability of large computational power has paved the way for novel techniques to analyze and simulate textile and parachute dynamics that can be incorporated into the design process~\cite{Robust,Val_far,comp_disk,immers,compliant}. Models under development range in scope from those applicable to the micro-scale permeability of textiles on a fiber-to-fiber basis~\cite{8,9,10,PUMA}, to parachute-relevant changes in the structure of textiles on a meso-scale tow-to-tow basis~\cite{Hearley1}, to full-scale and macro-scale fluid-structure interaction (FSI) simulations~\cite{1,12,13,14,15,Asad2022_1,Asad2022_2,Rabinovitch2022,pantano2022_1,pantano2022_2,BOUSTANI2022107596}. Models that reflect the connection between fabric porosity, micro-structure, and canopy mechanics, drag, and their effect to parachute performance are sought. Experimental data at relevant scales is needed to inform and validate these models. 

With advances in Scanning Electron Microscopy (SEM), X-ray Micro-Computed Tomography ($\mu$-CT), and high-resolution photography, imaging down to sub-micron scales is now possible. Techniques of this nature that resolve tows and fibers have already been proven to great effect for imaging composites and complex 3-D woven materials~\cite{Czabaj,Utah,ArchPerm,Through,FilDist,FranTrack1,CompCut,OGCentroid,Orient,Vol_Seggy,NARESH2020108553,SONG2021109436}. Quantities of interest such as fiber orientations and fiber-to-fiber spacing~\cite{Czabaj,Utah,FilDist} and tow centroid locations and areas~\cite{FranTrack1,OGCentroid} under load can be captured. Previous works have analyzed small sub-samples of data using tailored image processing and manual segmentation tools. Extracting quantitative and statistical information that best describes textile architectures from noisy 3-D image reconstructions requires semi-automated tools such as the ones developed for this work. 

This work is devoted to the characterization of meso- and micro-scale structures of parachute textiles and their evolution when the material is subjected to radial in-plane loads. Two textiles used in actual parachute systems were studied, MIL-C-7020H Type III~\cite{7020H} and MIL-C-44378(GL) Type II~\cite{44378}. They were chosen for having different properties (e.g., porosity and permeability), and for featuring distinct micro structural characteristics, such that the effects of fiber packing and arrangement on the overall tensile response could be investigated. An in-plane tensile testing apparatus compatible with a laboratory $\mu$-CT instrument was designed to image the 3-D structure of the fabrics at varying load conditions. Scanned regions were segmented and analyzed using an in-house data analysis framework that combines deep learning-aided segmentation and semi-automated extraction of quantities of interest, initially documented by the authors in~\cite{Phillippe2023}. The data processing methodology tracks the centroid locations and dimensions of each tow in the material, and outputs local and global statistics of quantities of interest, such as areal porosity, crimp angle, and tow cross-sectional dimensions along with their associated strains. The experimental setup, textiles, and data analysis framework are described in Section~\ref{sec:methods}. Results are presented in Section~\ref{sec:results} paired with a discussion of the implications of material design and construction on the textile behavior. This is followed by results and discussion of a geometric relation proposed by Payne in 1978~\cite{payne_1978} used in tandem with observed strain data to replicate experimental areal porosity results from the test campaign.

\section{Materials and methods \label{sec:methods}}

\subsection{Textile characteristics}
\begin{table*}[ht!]
    \centering
    \caption{Selected textile construction and associated material properties}
    \begin{tabular}{c c c c }
    \\[-2ex]\hline \hline \\
        Material & Permeability [CFM] & Warp/Weft Tow Size [D] & Num Warp/Weft Fiber \\
        \hline
        7020 T.III & 100 - 160 & 40 / 70 & 13 / 34 \\
        44378 T.II & 30 - 50 & 30 / 30 & 10 / 10
    \end{tabular}
    \label{tab: properties}
\end{table*}

\begin{figure*}[h!]
	\centering
	\includegraphics[width=16 cm]{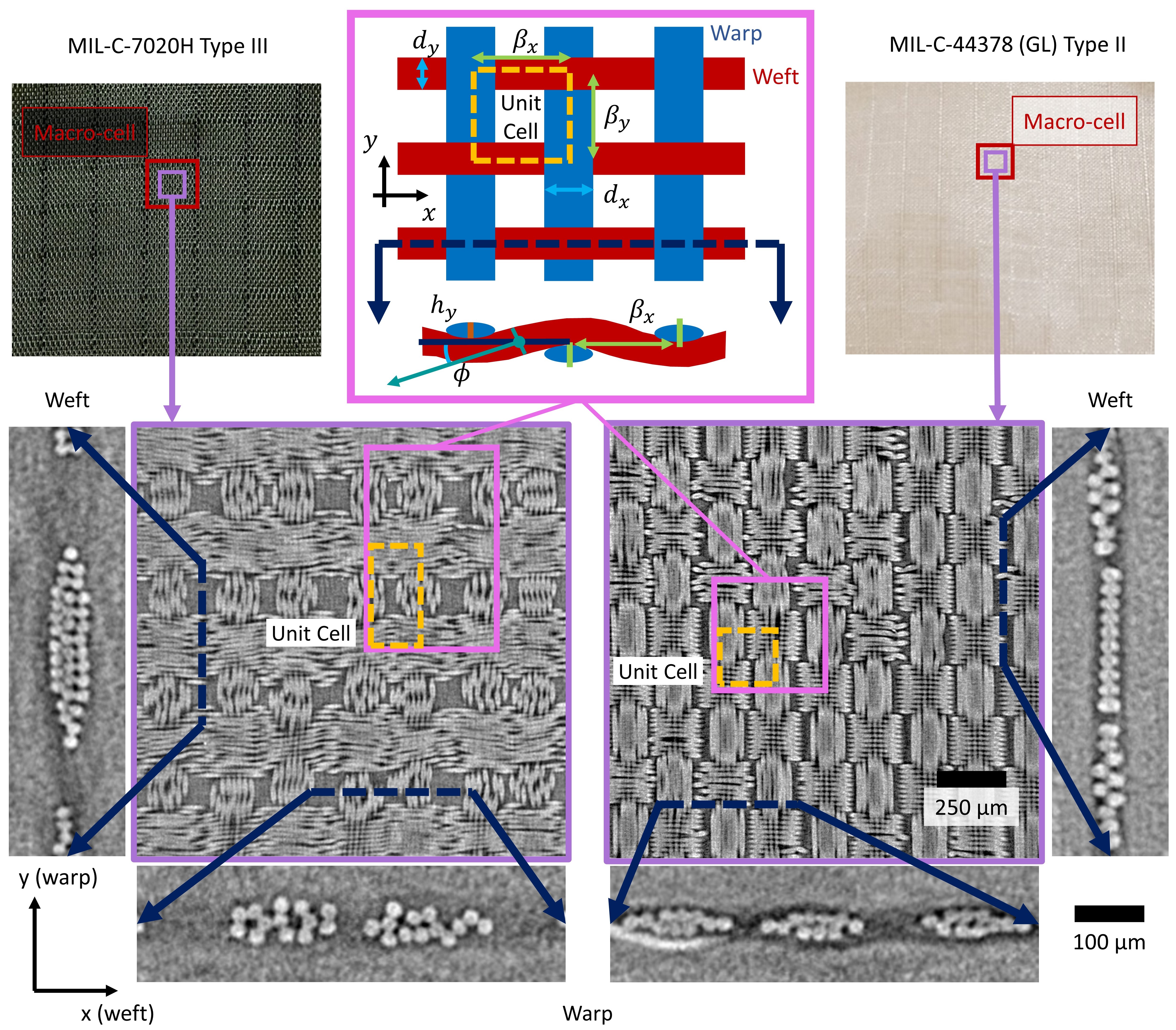}
	\caption{Images of MIL-C-7020H T.III (left) and MIL-C-44378(GL) T.II (right). Macro-cells (red), $\mu$-CT field-of-view (FOV, purple), and unit cell (yellow) highlighted. In-plane cross-sections of scanned region and out-of-plane cross-sections of distinct warp-weft tow fiber bundles are shown. Out-of-plane sectional views display tow peak-valley spacing, $\beta$ and crimp angle ($\phi$) definitions.}
 \label{fig: MyPayne}
\end{figure*}  
 The two textiles used in this work, MIL-C-7020H Type III and MIL-C-44378(GL) Type II, are referred to for brevity as 7020 T.III and 44378 T.II, respectively. The permeability specification is 100-160 CFM for 7020 T.III~\cite{7020H} and 30-50 CFM for 44378 T.II~\cite{ 44378}. Both are plain woven 1$\times$1 ripstop textiles. The plain weave is one of the three most common weave patterns alongside the varieties of twill and satin architectures~\cite{weaves}. It consists of perpendicular warp and weft tows that interlace in a simple over-under pattern, creating a grid-like structure. Micro-CT slices and cross sections of the two materials are shown in Fig.~\ref{fig: MyPayne}. Both materials are composed of nylon fibers but differ on the fiber diameters, the number of fibers that form the tows, and the tows spatial arrangement. The 7020 T.III textile is constructed with warp tows distributed in an alternating pattern of short and long inter-tow spacing and an even distribution of weft tows. In this material, warp and weft tows have different sized fibers with 13 fibers composing 40 D (Denier) warp tows and 34 fibers for 70 D weft tows. The 44378 T.II textile is also constructed with warp tows in an alternating pattern of short and long inter-tow spacing with evenly spaced 30 D weft tows, each containing 10 fibers. Fibers in weft tows are prevalently aligned into a single or double layer while warp tows feature the same 10 fibers compacted into an ellipsoid cross section, and thus have a smaller aspect ratio. Notably, the alternating pattern of warp tow spacing is less pronounced than the 7020 T.III. Material specifications for the two textiles are summarized in Table~\ref{tab: properties}. Note also the large difference in the size and pore count between the two materials. 44378 T.II has small pores, with characteristic pore widths between 1 and 10 $\mu$m, when compared to the larger ones present in the 7020 T.III, on the order of 10 to 100 $\mu$m. Both fabrics have thicker tows with double the fiber counts that break the material into distinct ripstop macro-cells visible to the naked eye in the textile photographs in Fig.~\ref{fig: MyPayne}. These structural tows are constructed from the same denier fibers as their counterpart warp and weft tows.

The schematic of a 2$\times$2 unit cell area (pink box) is shown in Fig.~\ref{fig: MyPayne}, including relevant weave geometric parameters. A single unit cell is denoted by the dashed yellow box. The in-plane or projected architecture is defined by the distance-between-centroids of parallel tows, $\beta$, and the tow projected diameter, $d$, also referred to as the tow width. The tow crimp angle, $\phi$, and the tow thickness, $h$, are shown in the schematic of the weave cross-section. The crimp angle is the angle of ascent of the tow centroid with respect to the centerline of the tow path at the point were the tow path intersects the centerline.

During weaving of plain woven textiles, the warp tows are pretensioned and held still while the weft tows are woven in between them. This results in higher crimp angles for weft tows. It is shown later in this work that the pretensioning contributes to unit cell anisotropic behavior.

\subsection{Tensile Tester and Experimental Methods}

\begin{figure}[h!]
\centering
\includegraphics[width=5.5 cm]{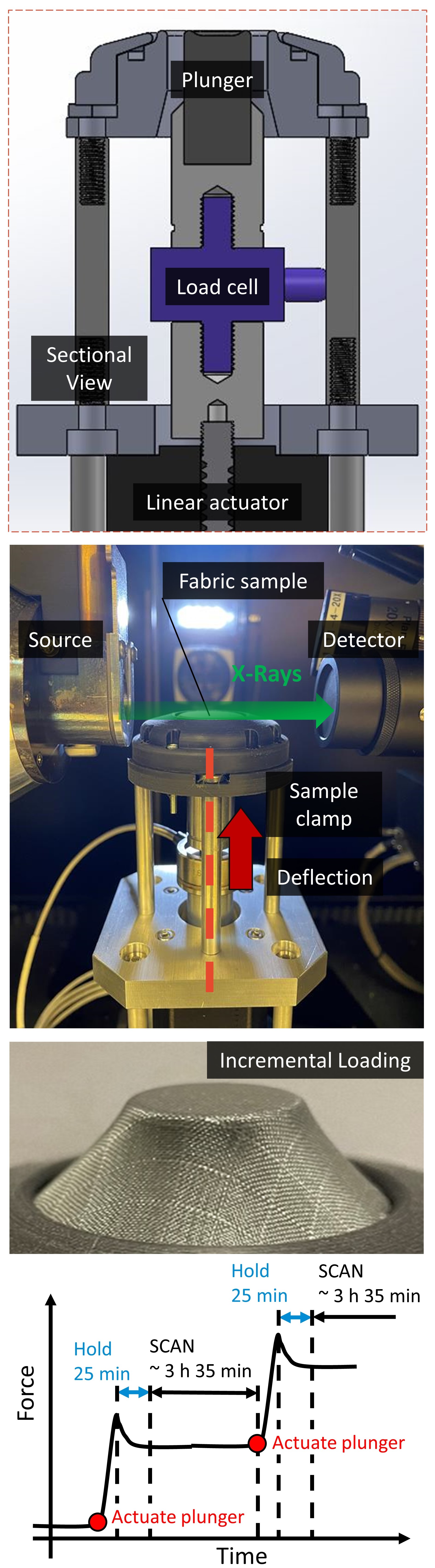}
\caption{Synopsis of tensile testing apparatus design, $\mu$-CT setup, and plunger testing procedure.}
\label{fig: TestProcedure}
\end{figure} 

A 2-D tensile tester was designed for in situ operation in an Xradia MicroCT scanner at The Beckman Institute at University of Illinois at Urbana-Champaign. The tensile tester and experimental setup in the scanner hutch are shown in Fig.~\ref{fig: TestProcedure}. The tensile tester consists of a sample clamp and a plunger, both machined from nylon to minimize X-ray absorption. The plunger inserts into an aluminum base above an in-line high-accuracy load cell (LCFD-250, Omega, Omega Engineering Inc., Norwalk, CT, USA). The load cell can support a maximum load of 250 lbf ($\sim$1100 N) operating at 2 mV/V with a linearity of 0.2\%. The load cell is attached at its base to a linear actuator (LN17D2200-E06008-070EN, Lin Engineering, 16245 Vineyard Blvd, Morgan Hill, CA, USA) for displacement control. The 24 VDC, 2 A/Phase linear actuator with the associated stepper controller (R525P, Lin Engineering, 16245 Vineyard Blvd, Morgan Hill, CA, USA) is capable of ~4 $\mu$m steps and can support up to 120 lbf ($\sim$530 N) in sample load. A 3-D printed base allows the apparatus to be mounted on the stage of the $\mu$-CT machine.  All fabric samples were cut to identical dimensions using a laser cutter (Model FSL PS48 Pro Series, Full Spectrum Laser, Las Vegas, NV, USA). Fabric laser-cutting was found effective at preventing fiber fraying. Samples were mounted into the apparatus via a circular clamp. The small initial load imparted upon clamping ranged between 0 to 6 N and was recorded for each test before loading. The material was then scanned as clamped before a step-wise load-to-failure scheme (Fig.\ref{fig: TestProcedure}) was applied via incremental plunger deflections. This scheme consisted of raising the plunger by $\sim$2 mm and waiting 25 minutes for the load to relax to steady-state. The static load was recorded, and a full tomographic scan was acquired. Single radiographs were acquired at each increase in plunger deflection to track displacement. The displacement increment was selected to most effectively replicate the full load-displacement curves of~\cite{Phillippe2023} in a time-efficient manner. 

A 4$\times$ magnification lens on the $\mu$-CT scanner provided a resolution of 1.9 $\mu$m/pixel within a cylindrical field of view (FOV) of $\sim$3.7 mm diameter and height. Fabric samples were pre-cut to 55.9 mm-diameter circular coupons and manually positioned on the 12.2 mm-diameter circular plunger such that large structural ripstop tows were not contained in the FOV. The FOV location was chosen at the center of the plunger to avoid plunger edge effects. While ripstop tows are also relevant for the overall material behavior under load, the focus of this study is on the bulk component of the weave architecture. Conflicting requirements exist between high-resolution imaging and large imaged regions due to the finite number of pixels on the imaging detectors. In this work we prioritized high-resolution imaging for  accurate characterization of material micro-structure at the expense of statistical information on the textile large-scale spatial variability. The plunger size, imaging magnification, and FOV, were optimized such that the imaged region contained several randomly chosen unit cells unaffected by plunger edge effects or neighboring ripstop tows. These selected parameters allowed for local statistical analysis of the material behaviour under load, but not results representative of the degree of randomness throughout the textile. Repeated tests would be required for a statistically relevant quantification.

\subsection{Image processing and analysis}

\begin{figure}[h!]
\centering
\includegraphics[width=5.6 cm]{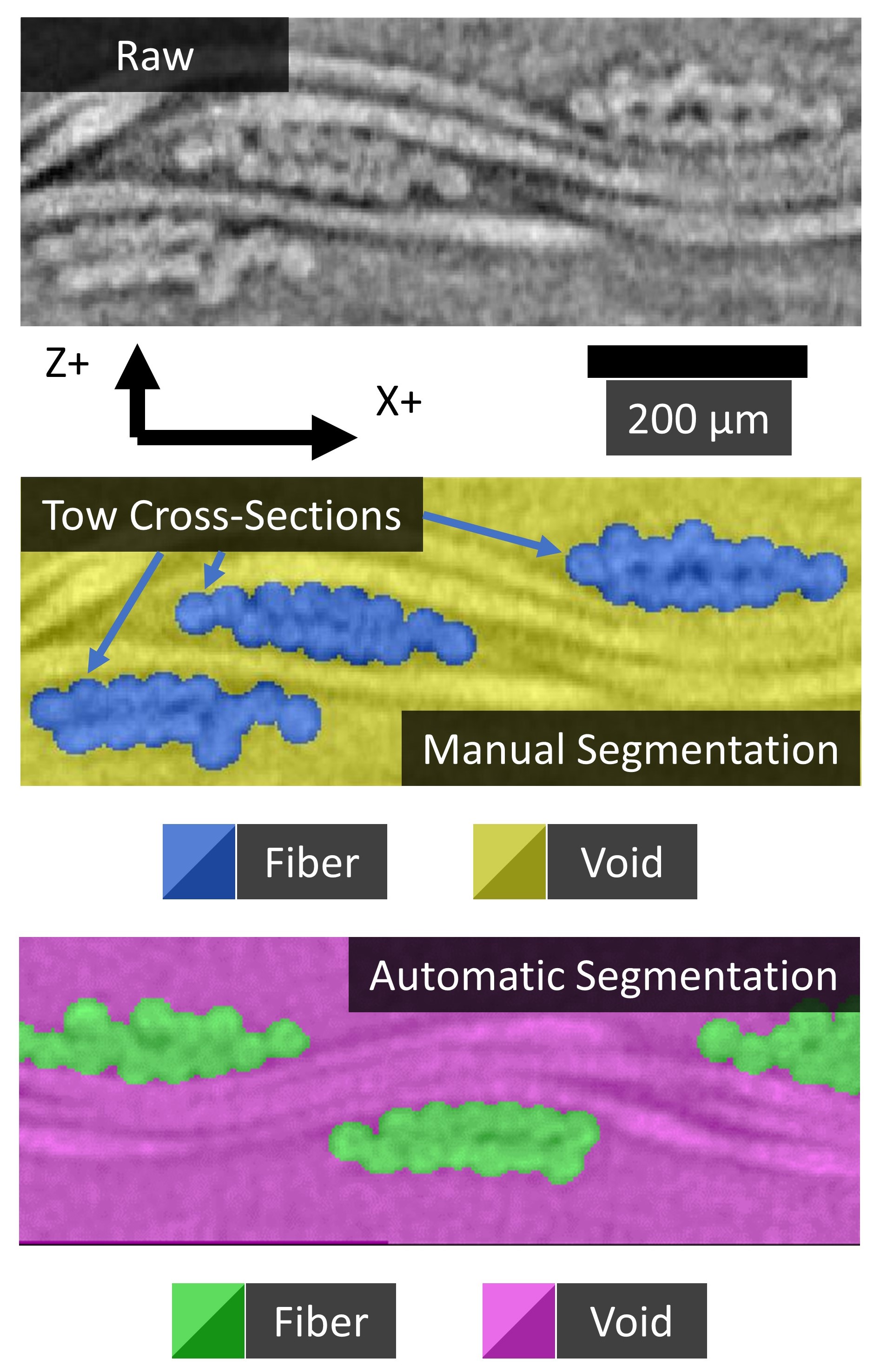}
\caption{Segmentation training: raw image (top), manual segmentation fed as training data (center), network segmentation post training (bottom).}
\label{fig: Training Image}
\end{figure}

\begin{figure}[h!]
\centering
\includegraphics[width=8.2 cm]{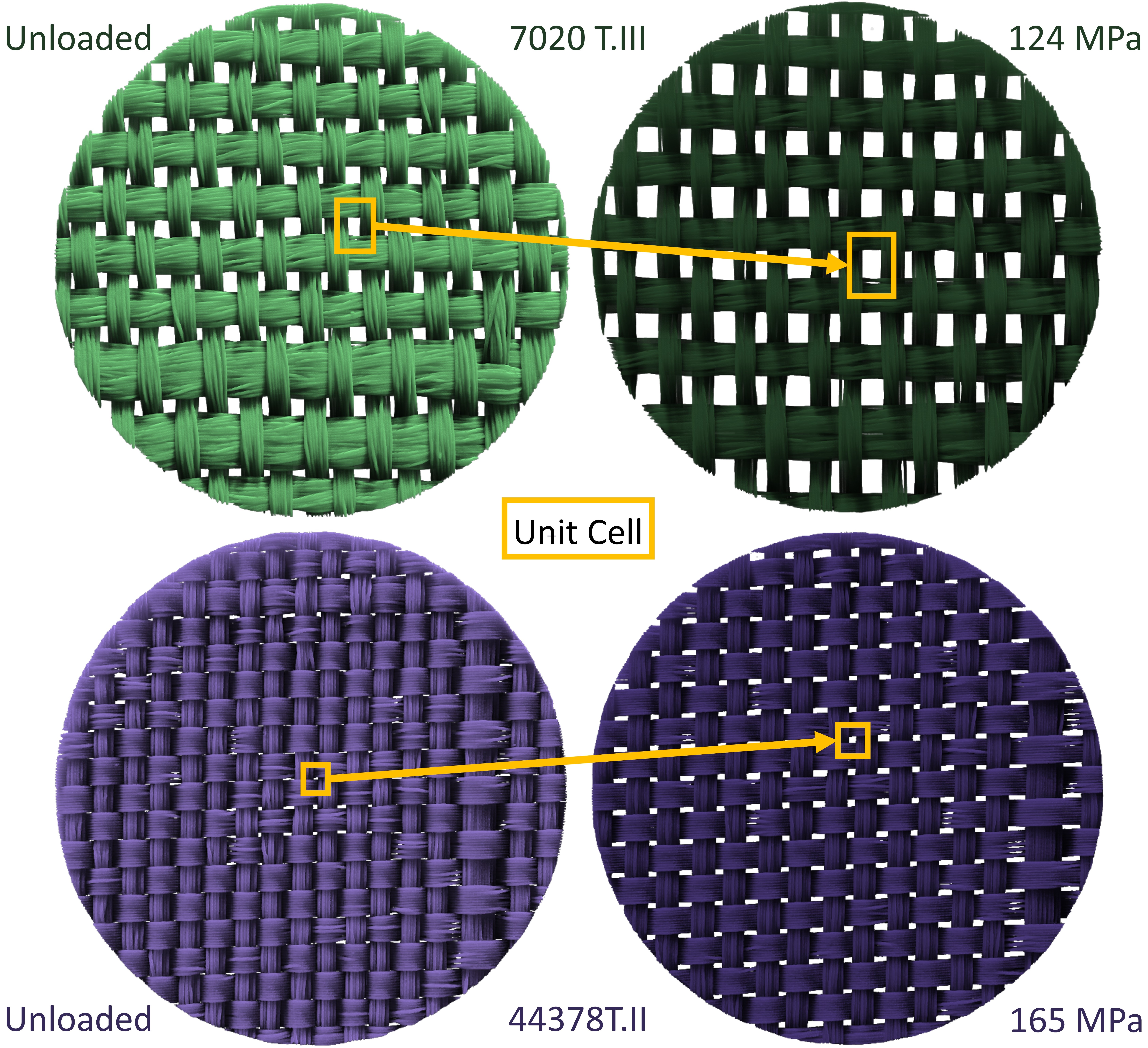}
\caption{3-D segmentation results: 7020 T.III (top) and 44378 T.II (bottom), for unloaded (left) and loaded (right) cases.}
\label{fig: Discs}
\end{figure}

Following reconstruction using the XRadia Bio MicroCT software, the output of each scan is a tiff stack of slices representing the 3-D volume of the scanned region. Reconstructed images were corrected for misalignment in the tensile apparatus with respect to the stack's horizontal $x$,$y$ plane using linear interpolation, and then segmented using the ORS Dragonfly image processing software~\cite{Dragonfly}. Semantic image segmentation was applied to separate warp and weft tows from each other and the background. A 3-D variant of a U-Net architecture~\cite{UNET} was used with a depth level of four and patch size of 64$\times$64$\times$64 pixels. The input training data consisted of manually segmented slices from the full datasets, three with the warp tows and three with the weft tows running normal to the plane of the slice. Modified versions of the manually-segmented datasets were automatically generated to enrich the pool of training data. These modifications consist of random rotations, transformations (stretching, shearing, etc), and noise additions within user-defined limits. The manually labelled pixels of one image experience the same rotation or transformation as the image they labelled, preserving the correct manual labeling of the modified versions of the images. The overlap between the labelling done by the trained network on these automatically generated image sets and the ground truth is quantified as the "validation loss". Neural network convergence to a validation loss of at least 6$\%$ was recorded before any application to the full datasets. The trained network was first evaluated on the same raw unsegmented images used for training and enriched with additional manual information when necessary. An example of this process is shown in Fig.~\ref{fig: Training Image}. Automated segmentation of the entire stack was done separately along the warp and weft tow axes. A final user guided visual verification was performed on the outputs from the entire warp and weft tow segmentation. Figure~\ref{fig: Discs} shows full 3-D segmentation of the scanned FOV in an unloaded and loaded case with a specific unit cell tracked between loads.


The segmented data were post-processed by a custom tracking code that uses a connected components algorithm~\cite{Phillippe2023, Phillippe2024} to identify individual tows and extract properties such as centroid location, width, and height along their path. The intersections between each tow centroid path with the plane of the weave were selected as reference locations to evaluate the tow crimp angle. The tracking code also resolves pore opening and changes to micro-scale fiber orientation by providing an in-depth 3-D view of tow behavior anywhere along the length of the tow.

\section{Results and Discussion \label{sec:results}}

\subsection{Principal Radial Stress}

  Membrane theory was applied to analyze the forces experienced by the sample, owing to the material thickness ($t^3$ = $\mathcal{O}$(E-12 mm)) being much smaller than the plunger diameter (12.2 mm). For the thickest sample the flexural modulus was estimated to be $D$ = $\mathcal{O}$(E-3 Pa). Since the macro-scale moment reactions experienced by the sample are proportional to the flexural modulus, they can be considered negligible. The force balance required to compute the sample radial tension is thus simplified. In addition, friction between the textile and the plunger was also considered negligible owing to the low coefficient of friction for both the nylon textile and nylon plunger~\cite{nylon1, nylon2}. Under these assumptions, the load induced on the sample can be translated to a total radial stress at the lip of the plunger using the setup geometry and textile cross-sectional area (cf. Supplement A; or go to the Supplemental Materials link that accompanies the electronic version of this article at http://arc.aiaa.org.). 

\begin{figure}[h!]
\centering
\includegraphics[width=6.4 cm]{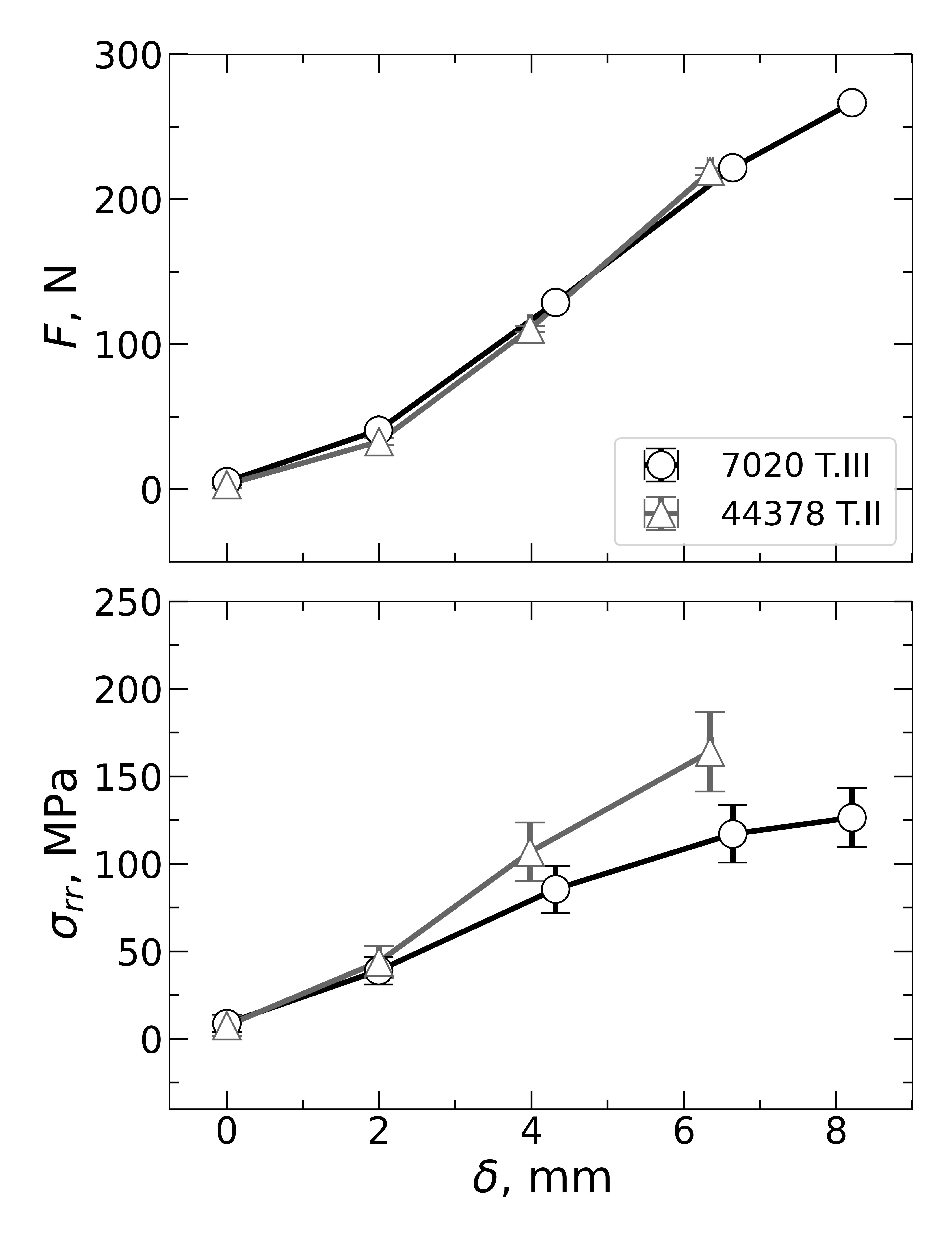}
\caption{Plunger load (top) and radial stress (bottom) as a function of deflection.\label{fig: Rad V Deflect}}
\end{figure} 

Figure~\ref{fig: Rad V Deflect} shows the force recorded on the plunger $F$ and the total radial stress $\sigma_{rr}$ as a function of plunger deflection for the two materials. At deflections beyond 2 mm,  44378 T.II displayed a higher stiffness. Both materials ruptured during the deflection step after the last recorded load. The different behavior of the two materials is attributed to the dissimilar fiber re-organization within tows as load increases, increases in tow spacing (which decreases sample cross-sectional area, $A$), and tow de-crimping, all of which are discussed in the following analysis. 

\subsection{Effect of Weaving Method on Crimp Angle Behavior}
\label{sec: crimp}

\begin{figure}[h!]
\centering
\includegraphics[width=6.4 cm]{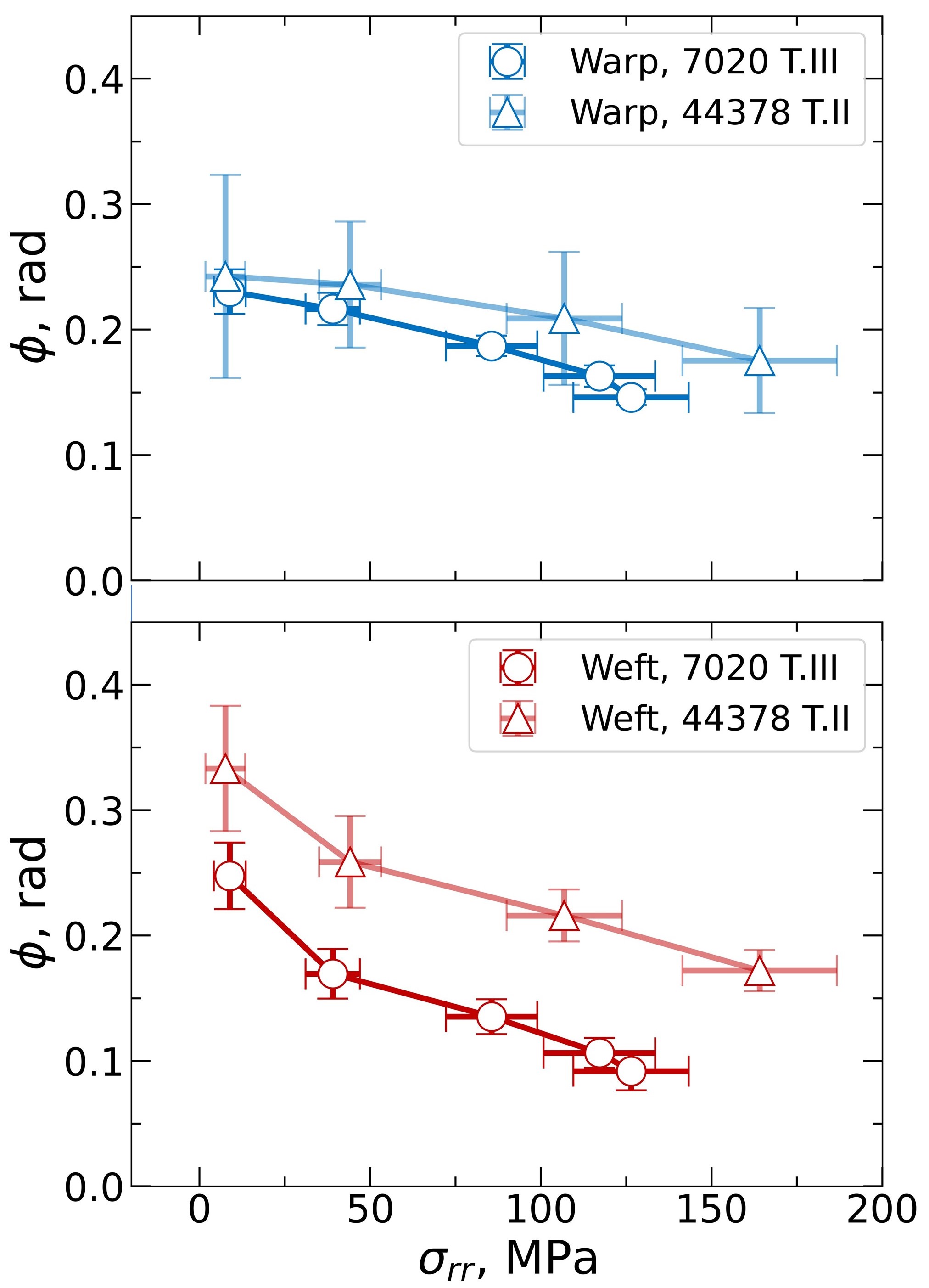}
\caption{Average crimp angle for increasing radial stress of warp (top) and weft (bottom) tows.}
\label{fig: Crimp Indiv1 V Load}
\end{figure}

Crimp angles were computed for all tows in the field of view by fitting the centroid data immediately surrounding the points where tows cross their respective material centerlines with a linear regression model. Field of view averaged crimp angles for both materials' warp and weft tows are shown in Fig.~\ref{fig: Crimp Indiv1 V Load}. Error bars represent the standard deviation of all crimp angles within a field of view, including the error of the linear fitting used to calculate the crimp angles. 


\begin{figure}[h!]
\centering
\includegraphics[width=6.8 cm]{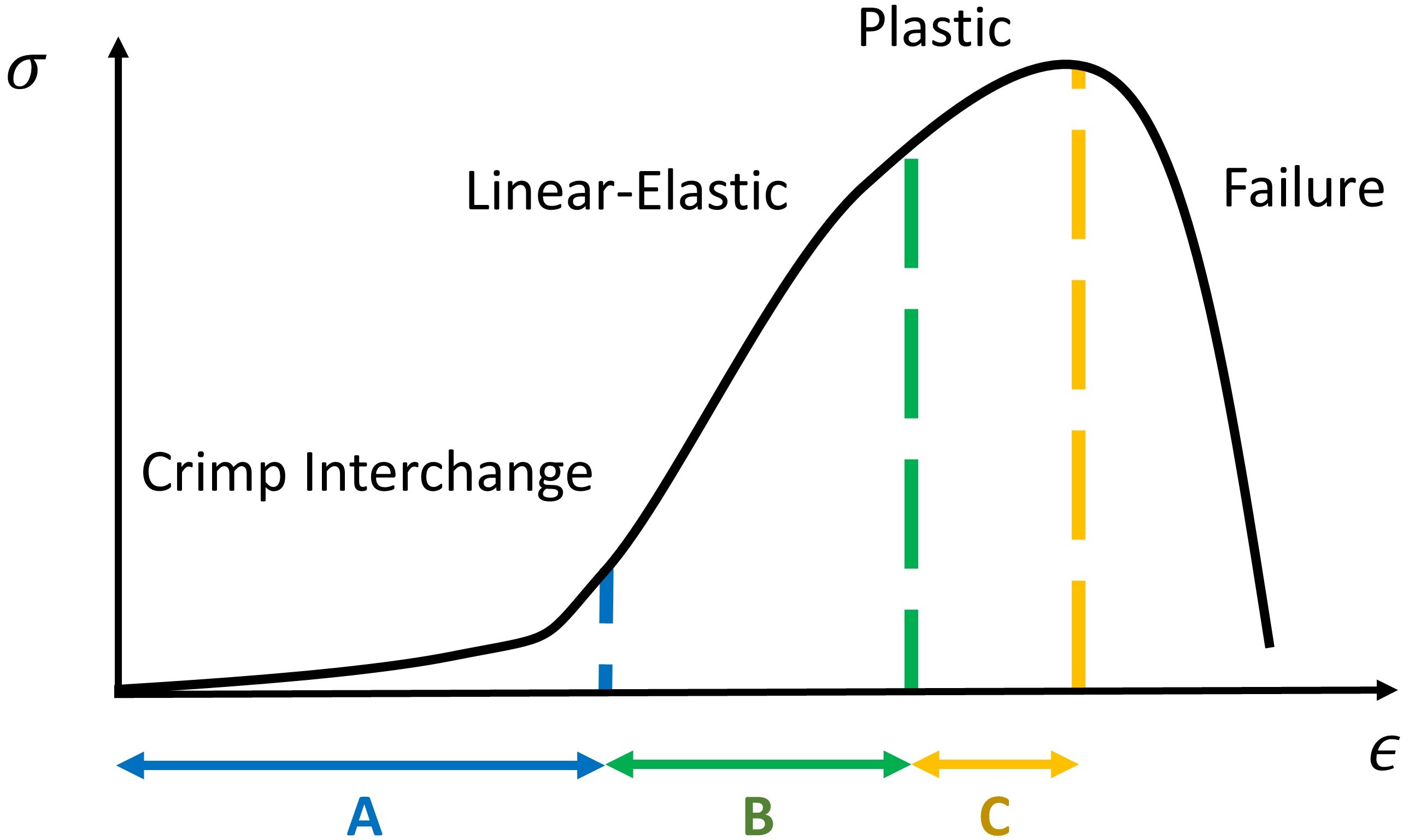}
\caption{Stress-strain schematic for general uni-axial tension on a textile with regions of behavior marked alphabetically.}
\label{fig: Regions}
\end{figure} 

\begin{figure}[h!]
\centering
\includegraphics[width=6.4 cm]{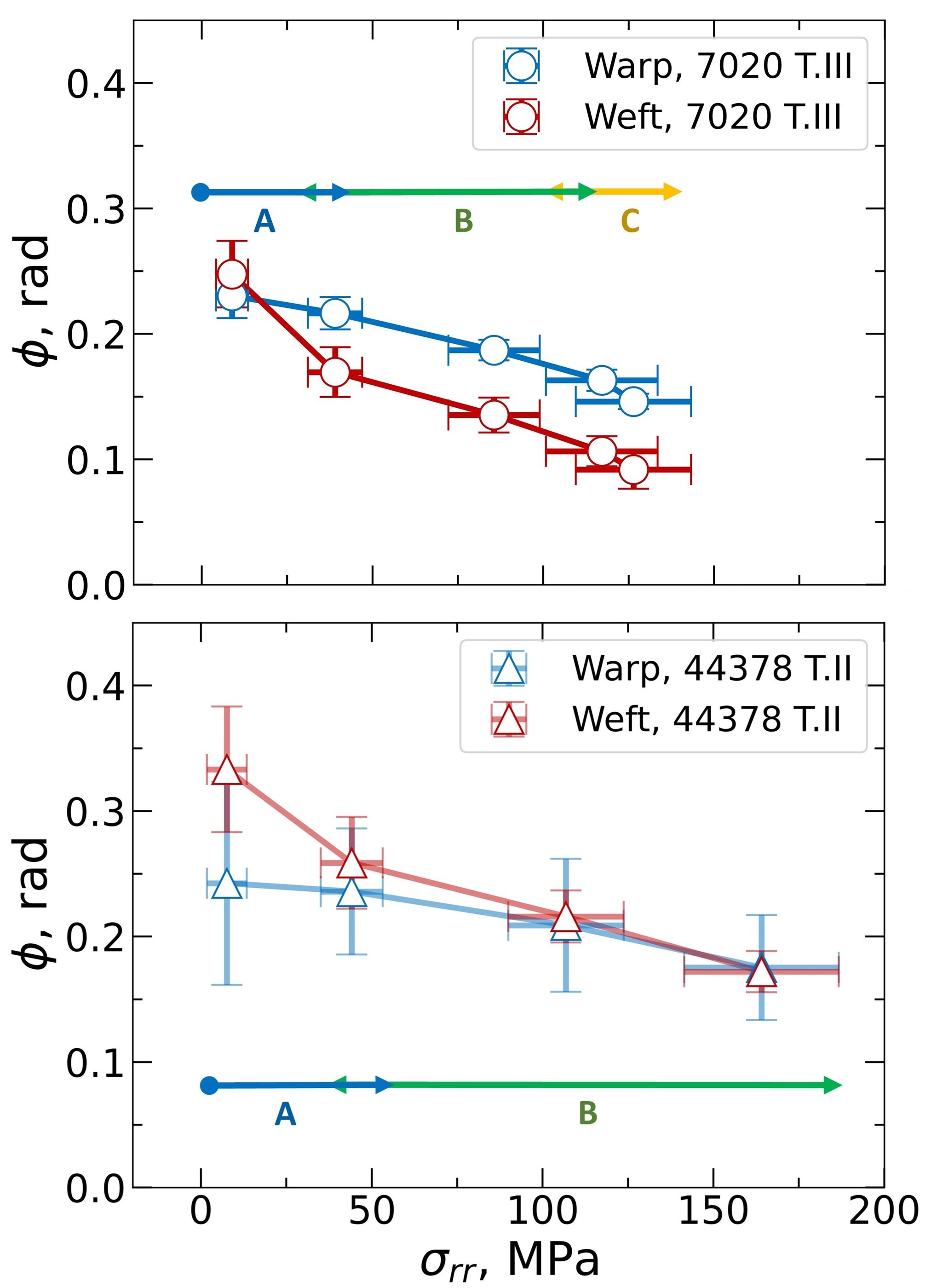}
\caption{Warp and weft average crimp angle comparisons for 7020 T.III (top) and 44378 T.II (bottom) with increasing radial stress, with stress-strain region correspondence.}
\label{fig: Crimp Indiv2 V Load}
\end{figure} 


A progressive tension-induced de-crimping is observed, which is more pronounced for weft than warp tows due to the pretension of warp tows during manufacturing. Warp and weft de-crimping with radial stress follows similar trends in both materials. Figure~\ref{fig: Regions} shows a typical stress-strain curve for loaded parachute textiles~\cite{ZHU20122021, POTLURI2007405, Hearley1, Hearley2} where four regions with different behavior are identified. The initial low stiffness region A, known as the "crimp interchange", is characterized by a nonlinear behavior that corresponds to tow de-crimping. Once de-crimping slows and the interlinked tows reach a "jamming" state, region B marks a linear-elastic regime in which de-crimping no longer dominates the macro-scale behavior of the textile. Region C corresponds to the transition from linear-elastic to plastic behavior and ultimately to failure. These regions are identified in the crimp angle versus radial stress curves for each of the two fabrics analyzed in Fig.~\ref{fig: Crimp Indiv2 V Load}, with results presented by material to highlight the different interplay of warp and weft behavior. For both textiles the first load step is attributed to region A, where de-crimping dominates. The second to third load steps correspond to region B, with weft de-crimping slowing in its jamming state. The radial stress versus plunger deflection curve (Fig.~\ref{fig: Rad V Deflect}) for the less stiff 7020 T.III textile showed a plateauing trend immediately before failure. Those last two test points for this material in Fig.~\ref{fig: Crimp Indiv2 V Load} indicate a plastic behavior transition to region C. No plastic regime or plateau of the stress vs deflection curve was captured for the 44378 T.II fabric before failure for the deflections tested; smaller increments of plunger displacement would be required. The 44378 T.II fabric is believed to be in the crimp interchange and linear-elastic regimes at all tested load conditions. Warp and weft tows, having equal construction and cross-sectional area in this material (Table~\ref{tab: properties}), show dissimilar initial crimp angles due to pretension of warp tows during manufacturing and thus, higher crimp angles of weft tows at zero-load. As the fabric is loaded, the faster decrimping of weft tows leads to converging crimp angles in both directions. The 7020 T.III textile has thicker, and therefore more rigid, weft tows compared with warp tows, which leads to similar crimp angles in both directions at zero-load despite the warp tow pretension. As load increases, weft tows decrimp faster leading to flatter weft tows with smaller crimp angles than warp ones. 

\subsection{Effect of Construction on Material Strain Behavior}
\label{sec: strain maps}

The local unit cell strains were analyzed by tracking the average spacing between consecutive tow centroids, $\beta$, along evenly spaced subdivisions of their projected length. The unloaded unit cell width, $\beta_{0}$, and that of the loaded case were used to compute the unit cell strain, $\epsilon_i$, along the $i$ principal direction via Eq.~\ref{Eq.strain}, where $i=x$ refers to strains along weft tows and $i=y$ along warp tows. Tows and corresponding principal axes at all loading conditions remained aligned, with maximum deviations of less than 3.5 degrees. The error on the unit cell strain calculation due to this misalignment is below the measurement uncertainty.

\begin{equation}\label{Eq.strain}
    \epsilon_i = (\beta_i - \beta_{i0})/\beta_{i0} 
\end{equation}

\begin{figure}[h!]
\centering
\includegraphics[width=7.6 cm]{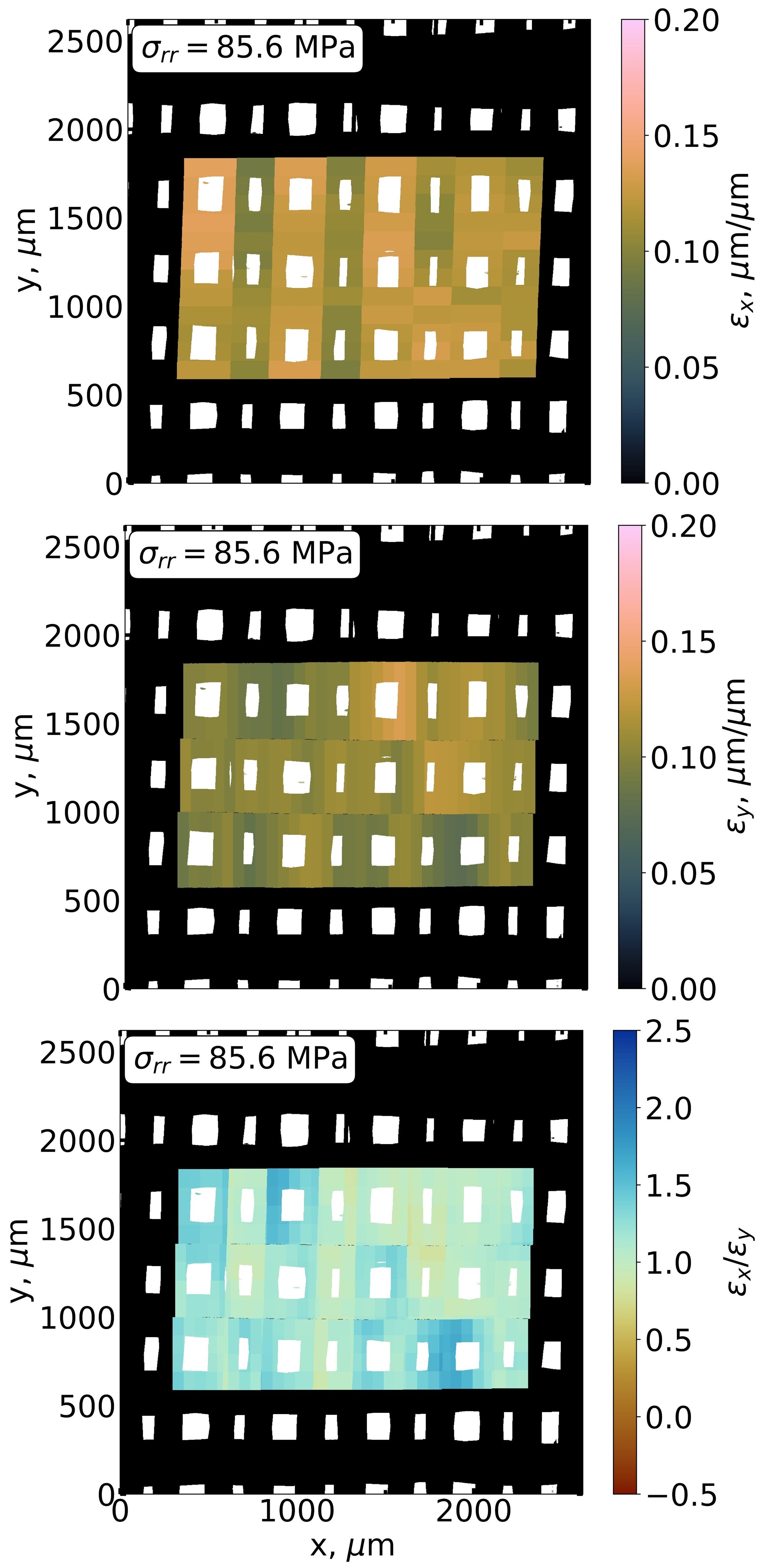}
\caption{Unit cell $x$-strain (top), $y$-strain (center), and $x$/$y$ strain ratio (bottom) maps for 7020 T.III at 85.6 MPa, projected onto the through-thickness view.}
\label{fig: map_process}
\end{figure} 

\begin{figure}[h!]
\centering
     \begin{subfigure}[b]{0.4\textwidth}
         \centering
         \includegraphics[width=\textwidth]{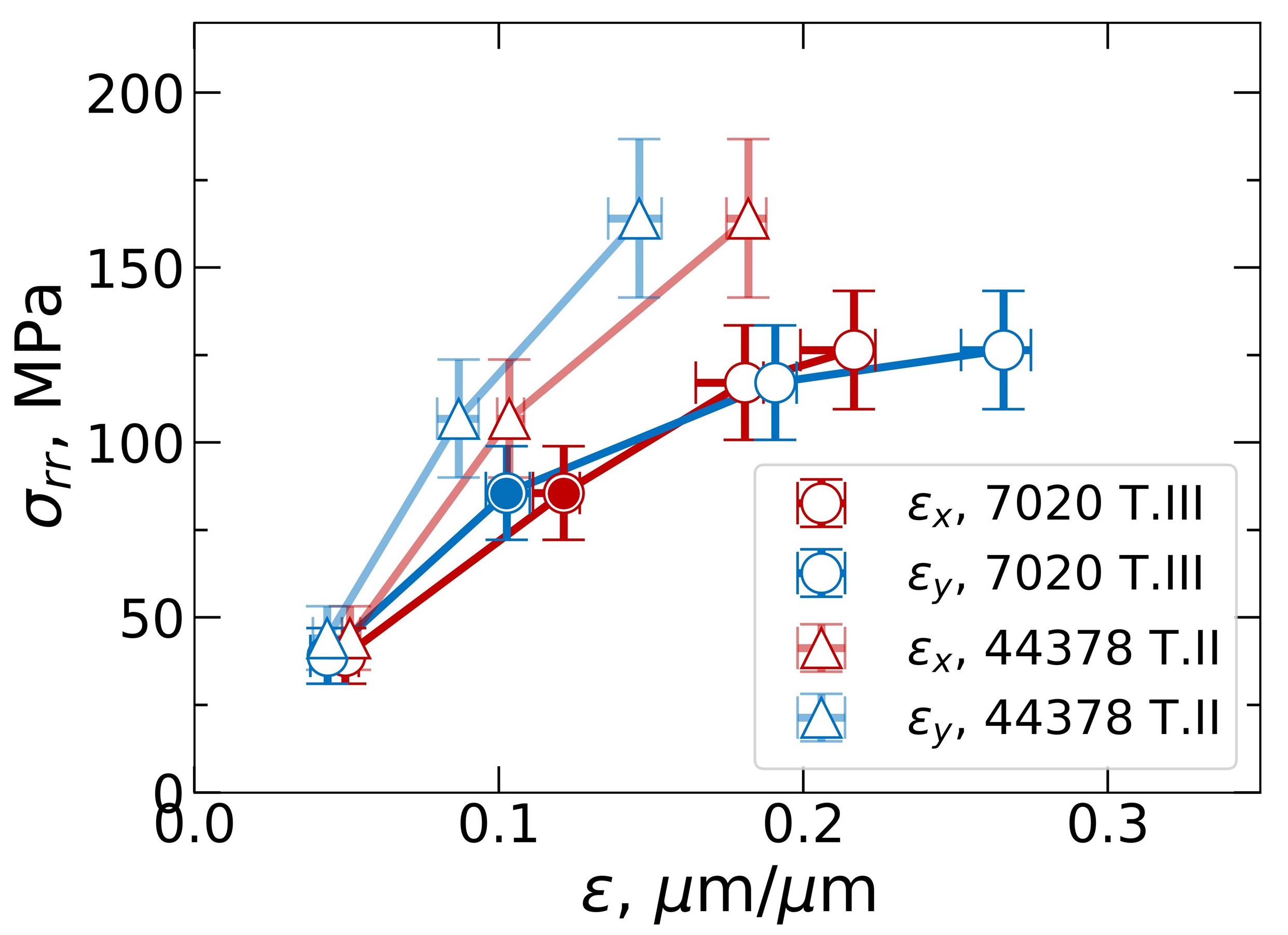}
         \label{fig: strain_plot}
     \end{subfigure}
     \vfill
     \begin{subfigure}[b]{0.42\textwidth}
         \centering
         \includegraphics[width=\textwidth]{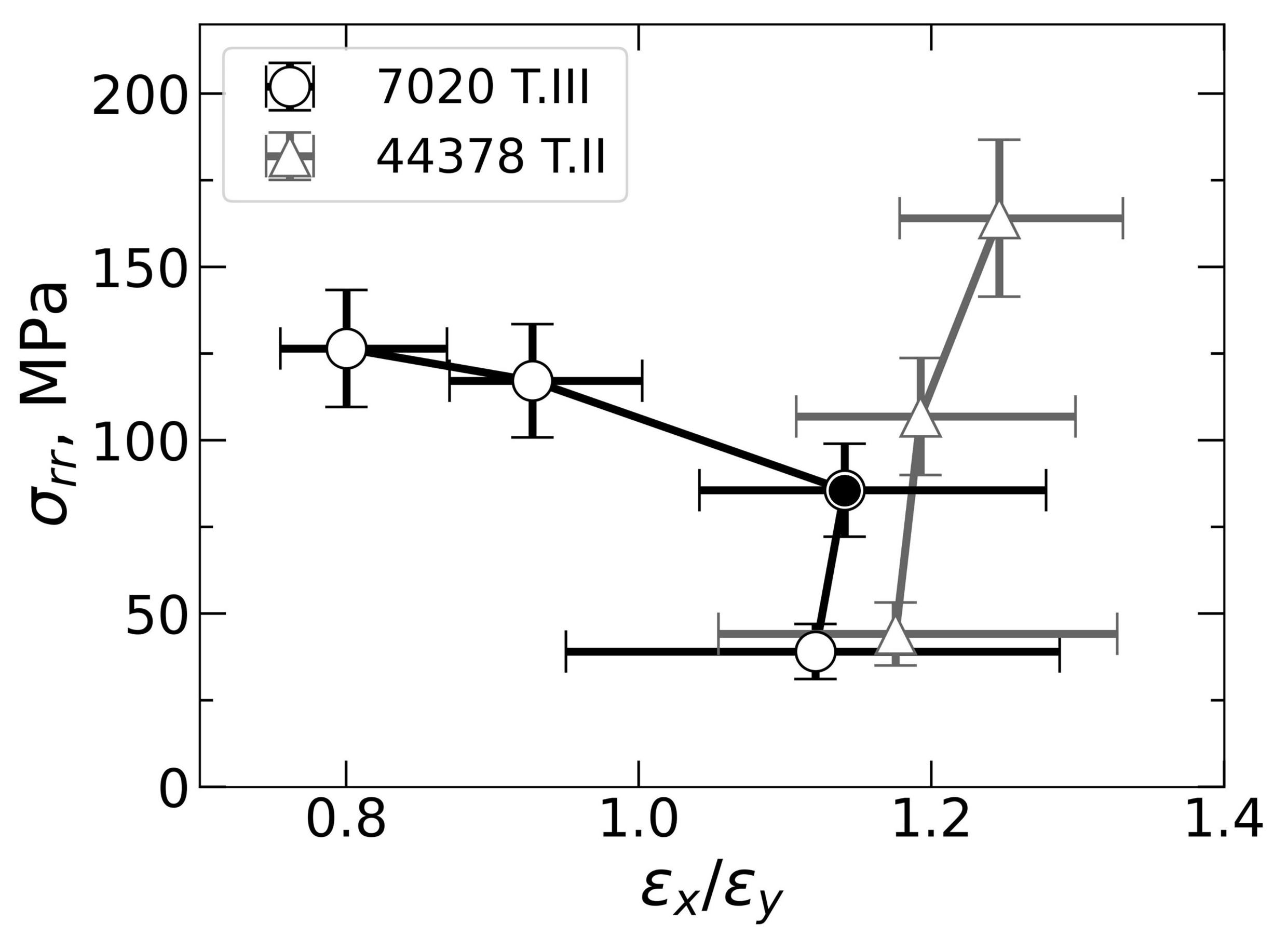}
         \label{fig: ratio_plot}
     \end{subfigure}
	\caption{Median-IQR statistics of principal axis strains for radial stress (top).  Median-IQR statistics for strain-ratio (bottom). Solid markers indicate median values corresponding to maps (Fig.\ref{fig: map_process}).}
	\label{fig: map_plots}
\end{figure}

Figure~\ref{fig: map_process} shows $x$ strain maps (top), $y$ strain maps (center), and their ratio (bottom), for the 7020 T.III material at a radial stress of 85.6 MPa during the second load case of plunger testing. For this specific fabric, $x$-strains fluctuate along weft tows, a pattern not identified in y-strains. The same alternating pattern is observed at all loads and it is not observable in the uniformly constructed 44378 T.II (cf. Supplement B; or go to the Supplemental Materials link that accompanies the electronic version of this article at  http://arc.aiaa.org.). The spatial distribution of strain is reflected in the line plots where each map has been reduced to its median and interquartile-range (IQR). The filled markers correspond to the maps displayed. Figure~\ref{fig: map_plots} condenses all principal direction strains for all radial stress conditions and both materials. Absolute values and relative increase of strain with radial stress are greater for the 7020 T.III (circular markers) than for the 44378 T.II (triangles). For a given material, the non-overlapping warp and weft strain evolution with load suggests an increase of anisotropy, also evident in the evolution with load of the strain ratio, in Fig.~\ref{fig: map_plots}. The greater initial decrimping of weft tows compared with warp tows for both materials (section~\ref{sec: crimp}) translates into greater unit cell $x$-strains compared with $y$-strains and, consequently,  $x-y$ strain ratios greater than one at the initial loads in all cases. $x$-strains continue to dominate at all loads for the 44378 T.II material, with strains along the weft direction ten to twenty percent higher than those along the warp. The trend is inverted for the 7020 T.III at the higher loads, which is hypothesized to be associated to the end of the linear-elastic regime and the transition to the plastic region (cf. Fig.~\ref{fig: Regions}). It is worth highlighting that 7020 T.III has less fibers in warp than weft tows, and that the two highest load points tested exceed the maximum load range estimated for an acreage gore in flight based on data from Mars 2020~\cite{1}. 

\begin{figure*}[htb!]
	\centering
	\includegraphics[width=16.5 cm]{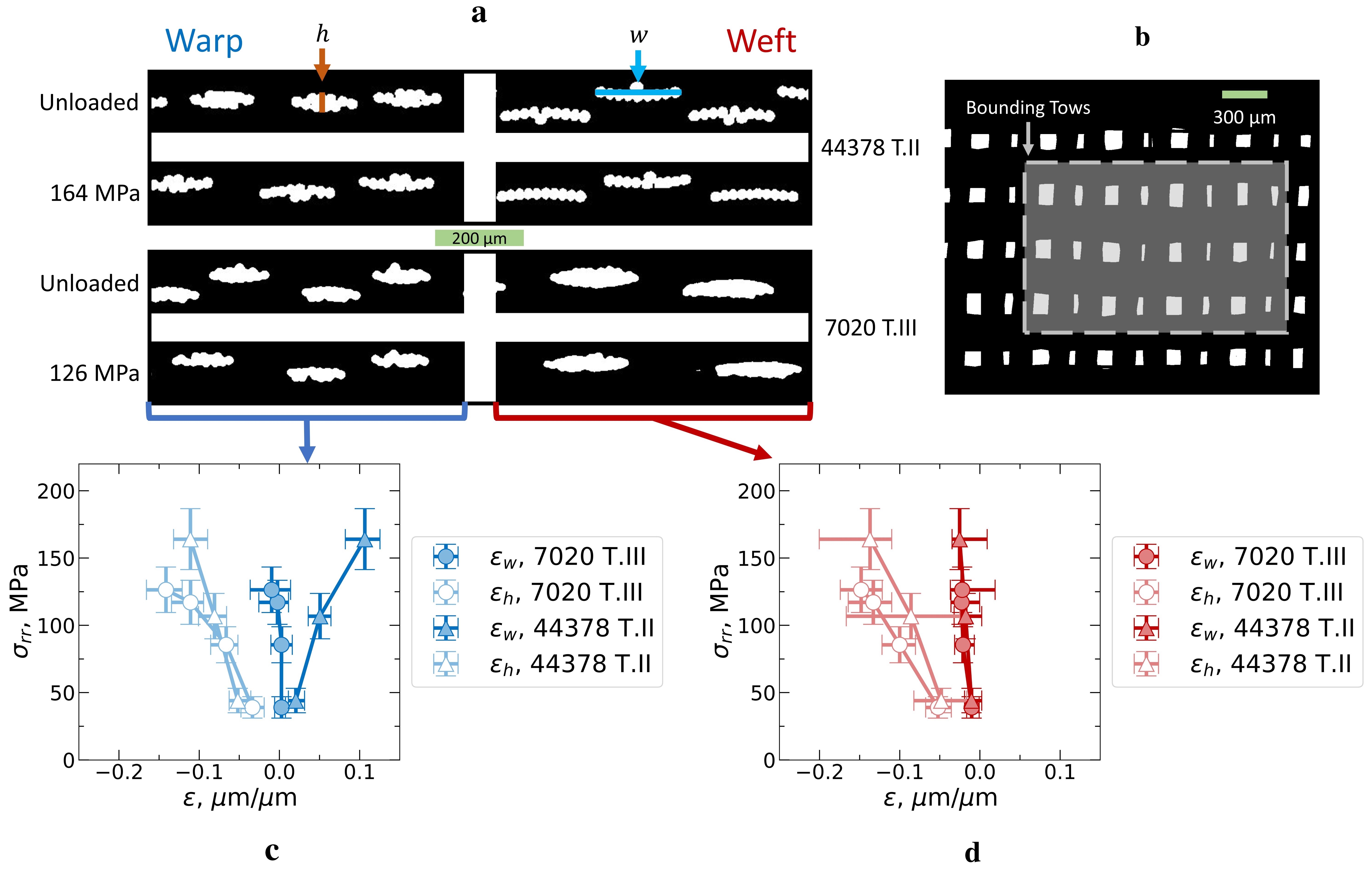}
	\caption{Tow cross section deformation under load for 7020 T.III and 44378 T.II (a). Region of interest considered (b). Strains of warp (c) and weft (d) tows' width and height for varying radial stress.}
 \label{fig: Tow Strain V Load}
\end{figure*}  

In addition to in-plane strains, the cross-sectional tow dimensions are also altered under applied load due to fiber reorganization. Tow cross-sections at unloaded and maximum load conditions are shown in Fig.~\ref{fig: Tow Strain V Load}a to qualitatively illustrate that fiber re-organization with load causes a decrease of the tow cross-section thickness, and therefore an increase of the tow cross-section aspect ratio. The change in shape was quantified by the change of each tow's in-plane width, $w$, and out-of-plane height, $h$. To ensure consistency, only tows within the central region of the field-of-view, highlighted in Fig.~\ref{fig: Tow Strain V Load}b, were considered. Warp and weft tows' width and height were averaged in unit cell subsections for each load and material, and strains were computed locally for width and height using the unloaded values as reference using the same definition as for principal axial strains. The median and IQR for each strain are represented for varying radial stress in Figs.~\ref{fig: Tow Strain V Load}c and ~\ref{fig: Tow Strain V Load}d for warp and weft tows, respectively. Warp and weft tows show a negligible variation of their width with load ($\epsilon_w$ close to 0) except for the warp tows of 44378 T.II that increase in width. In the tow cross section image this case displays fibers reorganizing away from a bundle with an increase in tow width. At the highest load they begin to resemble their more planar weft counterparts of the same construction. Variations of tow thickness are more significant, with tow thicknesses decreasing with increasing radial stress in all cases. 

Cross sectional and in-plane strains are not correlated, indicating that fibers re-organize leading to straightening of individual fiber paths but without playing a role in unit cell growth, which is dominated by tow de-crimping. Individual fiber linear-elastic or plastic behavior is not resolved in this work with the imaging resolution used.  

\subsection{Areal Porosity and Pore Size Distribution}

\begin{figure}[h!]
\centering
\includegraphics[width=8.2 cm]{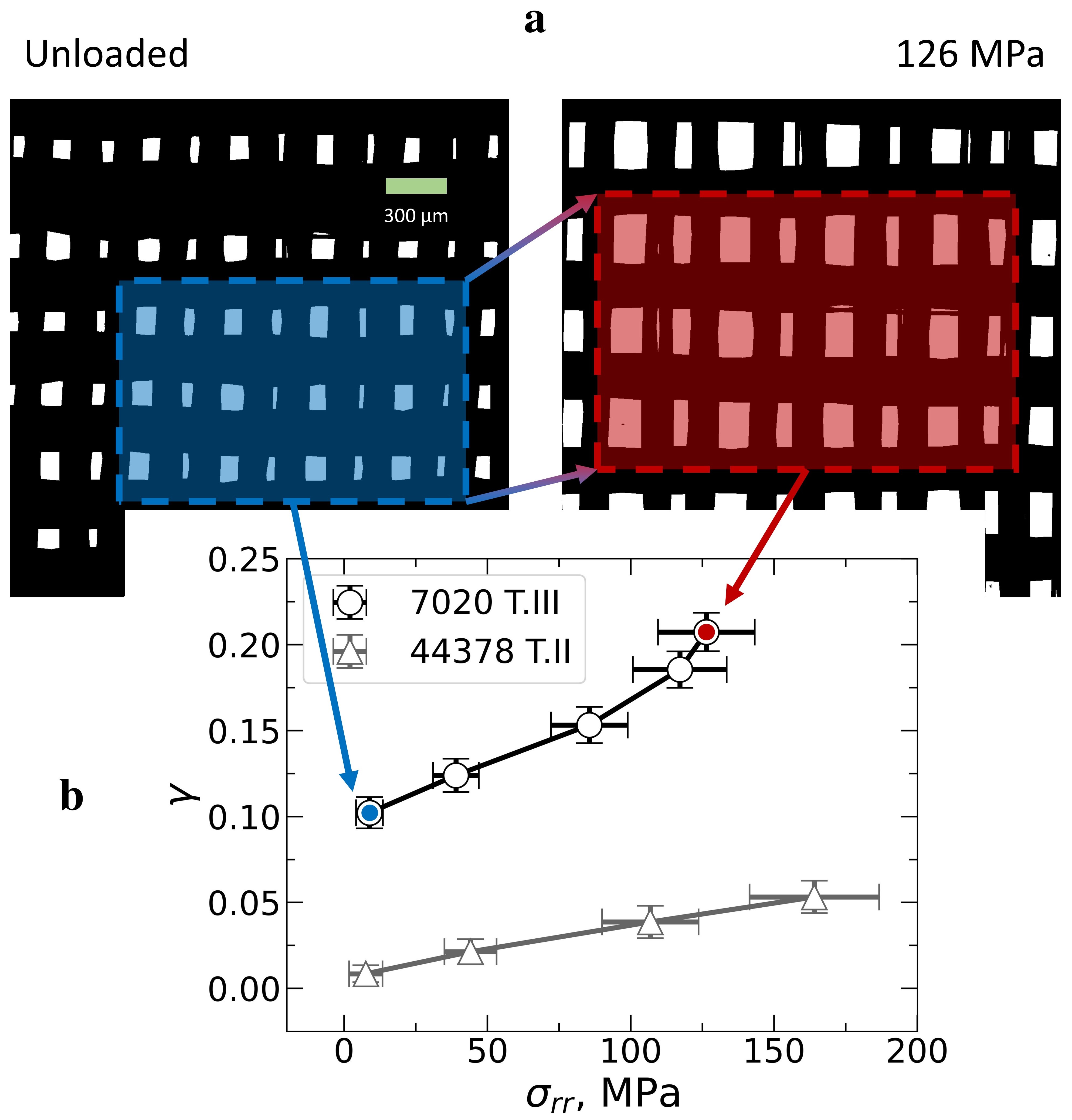}
\caption{Region of interest considered for pore ratio of 7020 T.III between unloaded and 126 MPa case (a). $\gamma$ for increasing radial stress (b). Solid markers indicate $\gamma$ of displayed region. \label{fig: Porosity}}
\end{figure}

Tow straightening and fiber re-organization with load effect the size and distribution of through-thickness pores and therefore directly impact the permeability of the textile by changing its porosity. This can be quantified by the "geometric" or areal porosity, $\gamma$, defined as the ratio of through-thickness pore area to total projected area. A region of textile bounded by four tows within the FOV at all loads was used to determine the areal porosity. The increase of $\gamma$ with load is reported in Fig~\ref{fig: Porosity} for both materials with error bars representing the resolution error. The 7020 T.III material experiences a greater growth of areal porosity with radial stress, and therefore a greater impact of load on permeability.

\begin{figure}[htb!]
\centering
\includegraphics[width=7.8 cm]{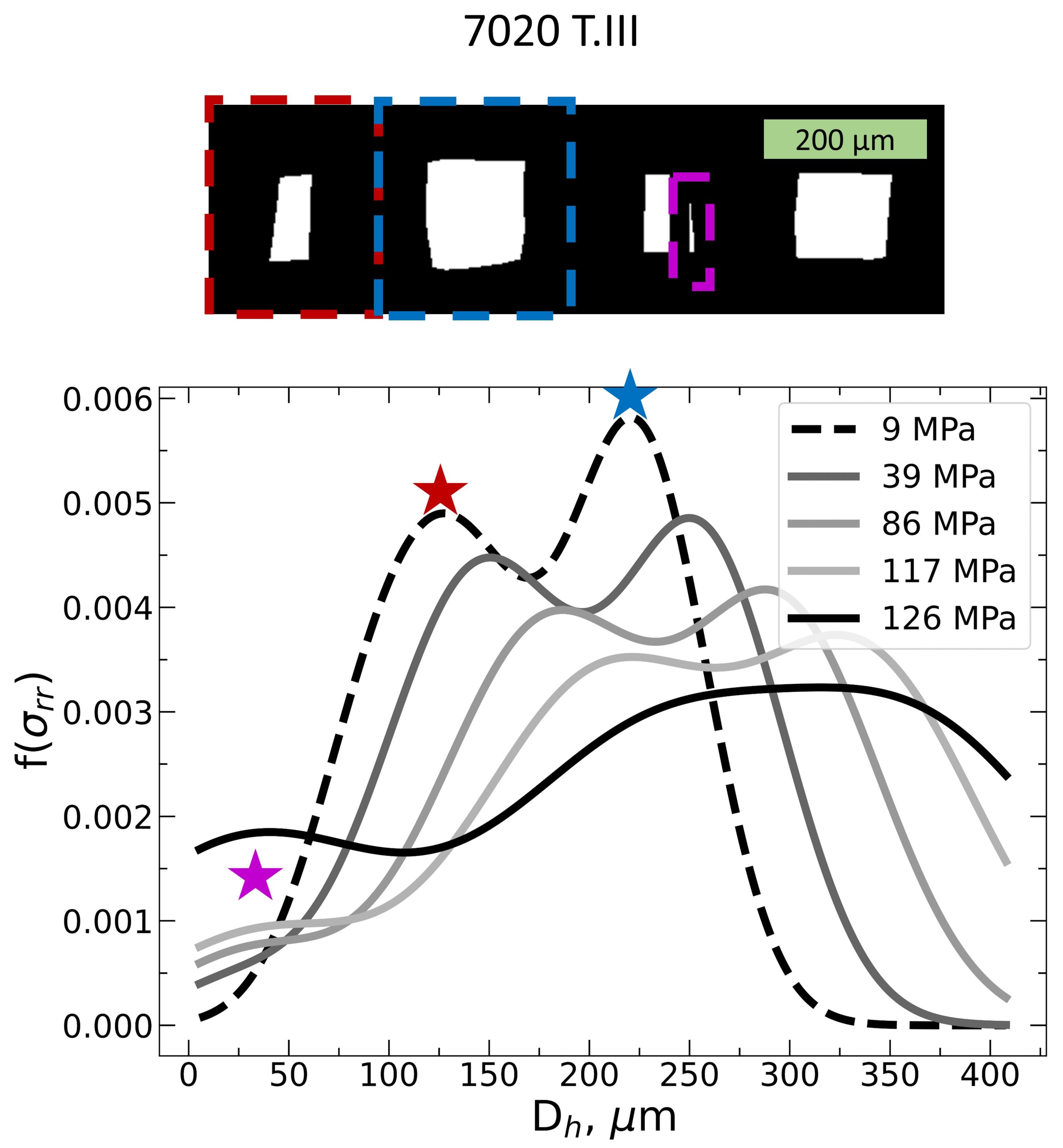}
\caption{Pore species of 7020 T.III: Inter-tow small (red), inter-tow large (blue), and intra-tow (magenta). Below, their respective impact on the probability density function (PDF) of pore sizes between scanned loads.\label{fig: Pores7020}}
\end{figure}

\begin{figure}[htb!]
\centering
\includegraphics[width=7.8 cm]{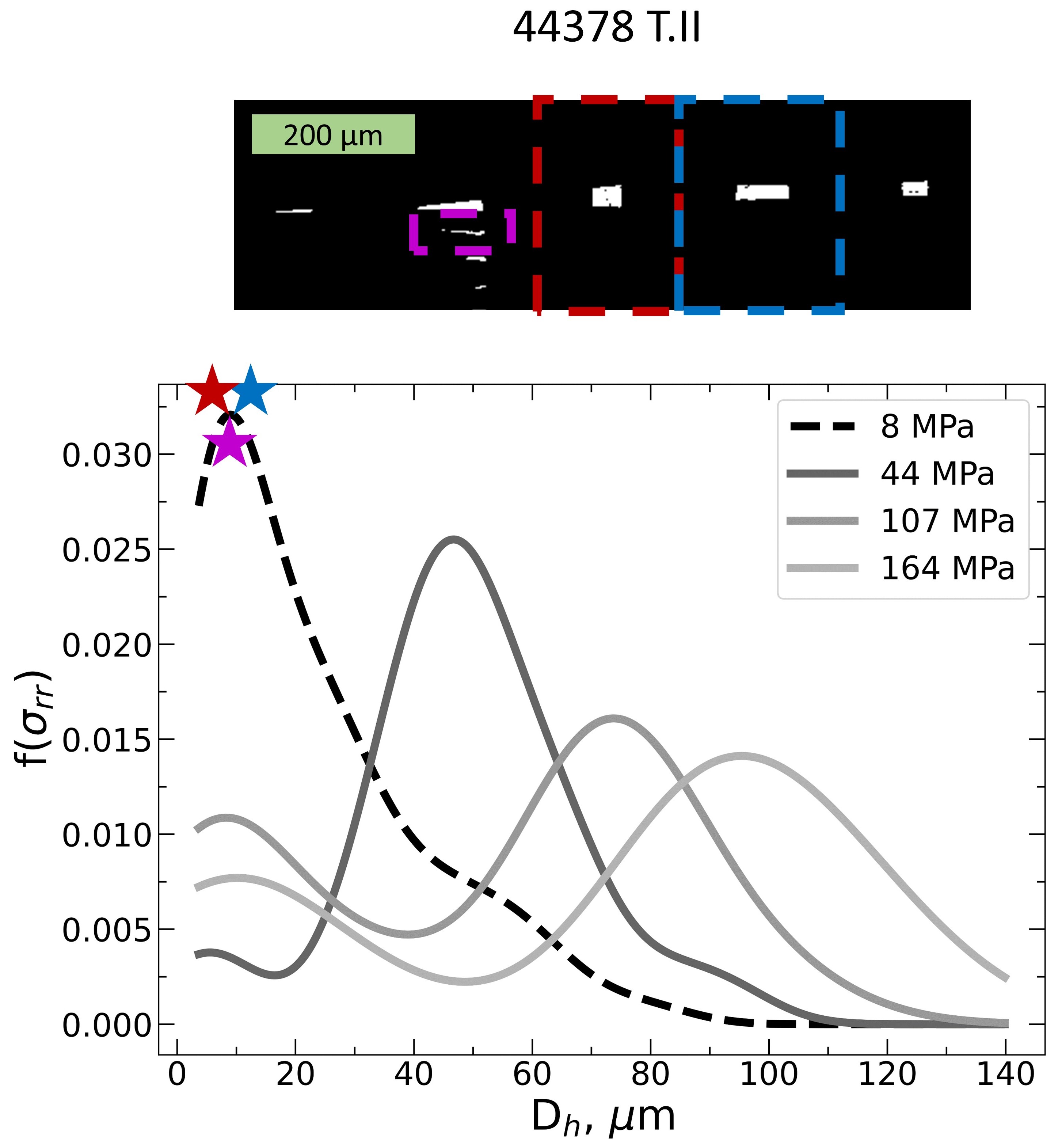}
\caption{Pore species of 44378 T.II: Inter-tow small (red), inter-tow large (blue), and intra-tow (magenta). Below, their respective impact on the  probability density function (PDF) of pore sizes between scanned loads.\label{fig: Pores44378}}
\end{figure}

In addition to the total areal porosity, the spatial variation of pore projected areas and the pores shape and size are important to the local flow features across the fabric and fabric effective permeability. Both materials present alternating small and large inter-tow pores bound by tow pairs (red and blue dashed regions in Figs.~\ref{fig: Pores7020} and~\ref{fig: Pores44378}), and smaller intra-tow pores (highlighted in magenta), which are a product of fiber size, number, and organization within a tow, and lack a predetermined distribution. The pore hydraulic diameter, $D_h$, is chosen here to evaluate the pore size and geometry's spatial variation.  


The probability distribution of pore hydraulic diameters, $D_h$, is shown in Figs.~\ref{fig: Pores7020} and~\ref{fig: Pores44378} for all load conditions for both materials. Kernel density estimates were used to build the PDFs due to the limited number of pores within each scan and are normalized to the range of observed pore sizes. The 7020 T.III textile shows initially two peaks corresponding to the inter-tow pores, with initial hydraulic diameters of 125 and 220 $\mu$m. As load increases their characteristic $D_h$ increases, as expected since the total areal porosity increases. However the $D_h$ distributions also become wider, with the two peaks becoming indistinguishable as the alternating pattern between inter-tow pore sizes becomes less marked. Intra-tow pores, with the smallest $D_h$, are non-existent in the region of interest at zero-load. As load increases, intra-tow pores remain low in relative number while also expanding their range. At the last load a large increase in their relative number is observed, indicating the appearance of multiple intra-tow pores. In the case of the 44378 T.II textile, of more homogeneous construction, inter-tow pores are indistinguishable from each other, no bi-modal distribution is observed. Inter-tow pores are smaller than in the 7020 T.III textile and present a lesser growth with load, also evident in Fig.~\ref{fig: Porosity}b. With increasing load the small intra-tow pores persist and grow in size while further more are formed. Overall, more intra-tow pores are present for this material relative to the number of inter-tow pores.

\subsection{Areal Porosity Estimation with Payne Relation}

Relations that capture the geometrical variation of textile properties with load are key to FSI numerical simulations of parachute systems~\cite{Asad2022_1, Asad2022_2, Rabinovitch2022,pantano2022_1, pantano2022_2,BOUSTANI2022107596}. A geometric model was derived by Payne~\cite{payne_1978} to calculate the areal porosity, $\gamma$, under strained conditions given the dimensions of a unit cell, the Poisson's ratios of the tow projected diameters (widths), and the strains experienced by both the unit cell and tow cross sections. Payne's formula uses the tow spacing, $\beta$, and tow widths, $d$, (cf. Fig.~\ref{fig: MyPayne}) to calculate $\gamma$ using Eq.~\ref{eqn: PayneSimple}.

\begin{equation}
    \gamma = \frac{(\beta_i - d_i)(\beta_j - d_j)}{\beta_i\beta_j}
    \label{eqn: PayneSimple}
\end{equation}

Here the projected pore area bounded by the tows, $(\beta_i - d_i)(\beta_j - d_j)$, is divided by the unit cell area, $\beta_i\beta_j$. Assuming the textile is composed of identical rectangular unit cells, the unit cell areal porosity is that of the material sample. If the unloaded unit cell geometry and the strains experienced by the unit cell during tension are known, it is possible to extend Eq.~\ref{eqn: PayneSimple} to estimate the areal porosity of a unit cell at a target strain condition. Assuming the strain of the overall textile is equal to that of each unit cell, the tow spacing in the $i$-direction, with $i=x,y$ for weft and warp, respectively,  becomes

\begin{equation}
    \beta_i = (1 + \epsilon_i)\beta_{i0}
    \label{eq:beta}
\end{equation}

where '$0$' denotes the initial zero-load state. The Poisson's ratio tying unit cell strain, $\epsilon_i$ to width strain, $\epsilon_{dj}$ can be determined as

\begin{equation}
    \nu_{dj} = -\frac{\epsilon_{dj}}{\epsilon_i} = -\frac{\Delta d_j\beta_{i0}}{\Delta\beta_{i}d_{j0}},
\end{equation}

\noindent where $\Delta d_j$ and $\Delta \beta_i$ respectively are the difference of the tow width and unit cell size from the zero-load state. Therefore the tow width, $d_j$, for a  strain perpendicular to the tow axis is given by

\begin{equation}
    d_j = (1 - \nu_{dj}\epsilon_{i})d_{j0}
    \label{eq:d}
\end{equation}

\noindent Substituting the strain-dependant forms of $\beta$ and $d$ (Eqs.~\ref{eq:beta},~\ref{eq:d}) into Payne's geometrical relation (Eq.~\ref{eqn: PayneSimple}) yields:



\begin{equation}
\begin{split}
\gamma &= \frac{(\beta_{i0}(1 + \epsilon_{i}) - d_{i0}(1 - \nu_{di}\epsilon_j))}{\beta_{i0}(1 + \epsilon_{i})\beta_{j0}(1 + \epsilon_j)} \\
&\quad \cdot ((\beta_{j0}(1 + \epsilon_j) - d_{j0}(1 - \nu_{dj}\epsilon_i)))
\end{split}
\label{eqn: PayneFull}
\end{equation}

\begin{figure}[!ht]
	\centering
	\includegraphics[width=8.0 cm]{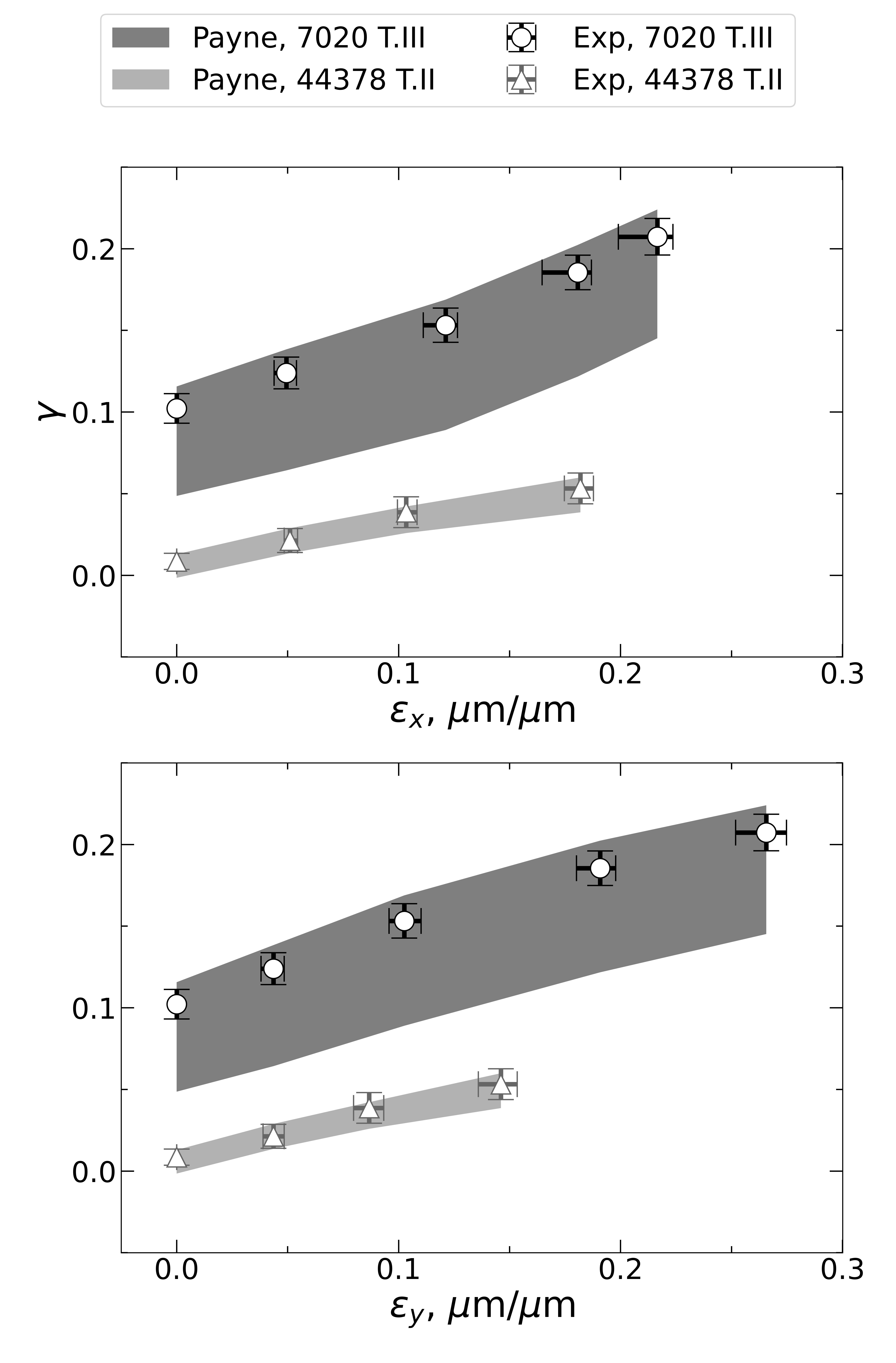}
	\caption{Comparison of measured and calculated (Eq. 6) areal porosity  as a function of measured directional strains.\label{fig: Payne}}
\end{figure}

Figure~\ref{fig: Payne} compares the experimental areal porosity results obtained as the ratio of through-thickness pore area to total projected area (cf. Fig.~\ref{fig: Porosity}b) with those calculated using Payne's relation. To evaluate Eq.~\ref{eqn: PayneFull} at each load condition, the values of each strain were evaluated locally and combined with the local unloaded unit cell and tow geometry to generate a distribution of areal porosities. The shaded regions in Fig.~\ref{fig: Payne} represent the interquartile range of each $\gamma$ distribution within the scanned field of view at a given load and consequent principal strains. Areal porosities versus $x$- and $y$- strains are shown separately for completeness albeit, not being independent variables.

The agreement observed between areal porosity obtained in these two different ways is expected as long as the fabric remains within the linear elastic loading regime, and indicates that purely geometrical relations can be used provided material strains are known. As discussed in section 3.2, that is the case for all load conditions except at the highest loads for the 7020 T.III material, where plastic deformation was identified. At the largest load points, the directly measured values of areal porosity are still within the IQR of the $\gamma$ values computed from Payne's relation, but deviations within the bounds of the error bars start to be observed. The experimental results stay near the upper end of the IQR for the 7020 T.III material. This is expected as $\gamma$ is tied to the relative surface area contribution of each of the measured unit cells. The larger inter-tow pores of the 7020 T.III constitute a larger percentage of the material surface area, despite being of equal number to the smaller pores. The larger pores therefore contribute more to the areal porosity of the bulk textile. In cases where $\epsilon_x$ and $\epsilon_y$ are either decoupled, isotropic, or parallel with load and no plastic deformation is occurring the simple geometric nature of Payne's relation implies a purely linear growth of $\gamma$ with strain. Should a material experience non-negligible and nonlinear changes in $d$ due to fiber reorganization, or there be the formation of very large intra-tow pores this linear assumption would fail. Intra-tow pores are not considered in Payne's relation, but the case of large non-linear changes in $d$ can be accounted for with experimental quantification of $\nu_d$.

\section{Conclusions}

This work presented a novel experimental and data-analysis methodology for in situ characterization under load of textile micro-mechanics. Two parachute textiles with dissimilar specifications were analyzed, namely  MIL-C-7020H Type III and MIL-C-44378(GL) Type II. The compiled datasets are key to derive constitutive model parameters to be incorporated in textile FSI solvers. Local changes in 3-D and projected 2-D textile topology with increasing load were presented. It was found that the anisotropic behavior of parachute textiles under planar load conditions is highly dependant on the method of manufacturing, induced pretension, and on the composition of warp and weft tows in both fiber number and size. Even in textiles with homogeneous tow composition, pretension during manufacturing leads to anisotropic behavior. Pretension of warp tows drives de-crimping with load of weft tows to outpace that of warp tows, leading to strains along the weft axis about 10 to 20\% larger than along the warp axis. The no-load ratio of crimp angle between the two orientations is modulated by the relative cross sectional area of the warp and weft tows. In the case of tows of equal construction no-load differences between crimp angles in both directions are larger due to the pretension than for the material with dissimilar warp and weft cross-sections. The relative magnitude of weft axial strain remains larger than warp axial strain for the material with homogeneous tow composition (44378 T.II) at all loads. However, the 7020 T.III fabric transitions to warp dominating strains at the two highest loads, which are believed to correspond to the end of the linear-elastic region with near-failure plastic deformation effects taking place. Tow width and thickness strains were also evaluated benefiting from the three-dimensionality of the micro-CT data. Fiber-reorganization with load was observed, but it does not play a major role for the selected nylon materials at the loads studied. Results were also presented on the increase of areal (or geometric) porosity with load for both textiles, as well as the spatial distribution of pore hydraulic diameters. Areal porosity is directly related to material permeability, and is key for FSI phenomena. A geometrical relation proposed by Payne~\cite{payne_1978} to estimate the areal porosity was proven effective at reproducing the values of areal porosity directly extracted from the $\mu$-CT scans for those loads in the linear-elastic regime. Given information is known about the distribution of 2-D axial unit-cell strains of a textile, this relation can be used instead of more computationally costly Finite Element Analysis of the material architecture to provide FSI simulations with areal porosity. 



\section{Data Availability}

The raw data required to reproduce these findings are available to download from [\url{https://doi.org/10.18126/82fx-q319}]~\cite{image_dataset}[dataset] at Materials Data Facility~\cite{mdf1, mdf2}. 

\section*{Funding Sources}

This work was supported by an Early Stage Innovations grant from NASA’s Space Technology Research Grants Program, NASA Grant No. 80NSSC21K0224. The authors would like to thank our research collaborator from NASA LaRC, Dr. Juan Cruz.

\section*{Acknowledgments}
The authors would also like to thank collaborators at Aviation Safety Resources, Scott Hilton with Aerial Delivery Solutions LLC, and undergraduate researchers Seongyong Hong and Alexa English for contributing their indispensable time, technical know-how, and textile samples.

We also acknowledge the Microscopy Suite at the Beckman Institute for Advanced Science and Technology.

\bibliography{sample}

\end{document}